\documentclass{article}

\usepackage{arxiv}

\usepackage[utf8]{inputenc} 
\usepackage[T1]{fontenc}    
\usepackage{hyperref}       
\usepackage{url}            
\usepackage{booktabs}       
\usepackage{amsfonts}       
\usepackage{nicefrac}       
\usepackage{microtype}      
\usepackage{lipsum}
\usepackage{graphicx}
\graphicspath{ {./images/} }

\title{A High Resolution Optimum 2D Coprime Planar Array}

\author{
 Kretika Goel \\
  Research Scholar\\
  SENSE\\
  IIT Delhi \\
  \texttt{Kretika.Goel@iddc.iitd.ac.in} \\
   \And
 Monika Aggarwal \\
  Professor\\
  CARE\\
  IIT DELHI \\
  \texttt{maggarwal@care.iitd.ernet.in} \\
  \And
 Subrat Kar \\
  Professor\\
  Dept of Electrical Engg\\
  IIT Delhi \\
  \texttt{subrat@ee.iitd.ac.in}\\
}

\begin{document}
\maketitle
\begin{abstract}
Designing a new class of rectangular two-dimensional sparse array to enhance the signal resolving capabilities with a limited number of sensors has always been a challenge. We explore the non-uniformity of the sparse arrays to enhance the Degrees of Freedom (DOF) by considering the under-determined cases using the concept of the virtual array. In this paper, we propose a two-dimensional novel sparse array configurations to estimate both azimuth and elevation angle of arrival of the signal. We propose a new class of rectangular two-dimensional arrays with sensors on a plane, whose difference co-array can give rise to a virtual two-dimensional planar array with a much larger number of elements, leading to an increase in the fully augmentable range of the array. The difference co-array of the proposed rectangular array gives a contiguous dense structure which leads to a reduction in holes within the planar array as compared to the traditional methods of CPA geometry in which there are many holes present in the outer dimension of the array structure as well as within the array as well. Moreover, we know that the angular resolution of an array is highly dependent on its beamwidth which itself is inversely proportional to array size or effective array aperture. Hence by increasing the array dimension a higher resolution can be achieved. The proposed arrays provide a higher DOF for the given number of physical sensors. Optimization is performed on the proposed geometry to maximize the directivity in the array steering look-up direction by suppressing the sidelobe levels to a greater extent. 
\end{abstract}


\section{Introduction}
Numerous applications of array signal processing exist in satellite communication, wireless communication, MIMO, the medical field, the military, radar, and sonar, among others. The fundamentals of array signal processing are as follows: processing the signal that strikes an antenna or array of antennas and then reproducing the received signal while suppressing noise or interference signals. The concept of signal spectrum distribution in different directions gives rise to spatial spectrum theory, which aids in determining DOA, or more precisely, spatial spectrum estimation, which serves as the foundation for array signal processing. DOA calculation is extremely useful in underwater acoustic networks \cite{partan2007survey},\cite{heidemann2006research}, wireless networking \cite{mao2007wireless}, localization techniques, tracking, navigation, etc. To improve DOA resolution, a critical method currently being used is to use an array of antennas rather than a single antenna, as was previously done, due to the sensor array's superior performance.

There are three distinct types of antenna arrays based on the location of the sensors within the array, namely linear arrays with sensors in a straight line. This is the simplest array design and is the most frequently used due to its simplicity of structure, but when extended array aperture and a high degree of freedom are required, a uniform linear array falls short. As a result, we introduced a sparse array, to improve the resolution. The sparse array's objective is to find the array configuration that produces the desired co-array with the fewest possible sensors. Another advantage of sparse arrays is that they contain fewer closely spaced elements, which results in less mutual coupling between the sensors.

\section{Types of Sparse Arrays}
\label{sec:headings}
Various classifications of non-uniform arrays include Minimum Redundancy Array(MRA) \cite{yangg2016new}, Minimum Hole Array (MHA) \cite{zhang2013sparsity}, , Nested Array \cite{zheng2020robust}, Coprime Array \cite{zhou2013decom}, etc. While most of the applications prefer uniform array structures but nonuniform geometries gain interest due to their efficiency in predicting the sources where the number of sources is more than the sensors available.
One such nonuniform geometry is MRA but there is no closed-form of expression available to obtain the sensor positions of an MRA. Then nested array was introduced but there comes the problem of mutual coupling in it hence coprime array was introduced.
Convention Coprime array consists of two subarrays having 2M and N elements respectively with inter-element spacing between them as Nd and Md where d is less than $\lambda/2$ where M and N are coprime numbers and d is inter-element spacing between the sensor elements. It is represented as follows:

\begin{equation}
P = \{Mnd | 0 \leq n \leq N -1\} \cup \{Nmd | 0 \leq m \leq 2M -1 \}  
\end{equation}
                    `
and it can resolve up to (2MN) sources with (2M+N-1) sensors. 

 To extend the research span from azimuth to elevation angle, a two-dimensional coprime array was investigated. Parallel arrays, non-parallel arrays, and planar arrays are all proposed as two-dimensional array structures. Two parallel antenna arrays are used, with subarray1 placed $d$ apart from subarray2, and the 2-D DOA estimation is performed first on the first subarray, then on the second subarray, and finally on the overlapped peaks in the array's phased spectrum. Angle pairing is a frequent occurrence in parallel arrays. Three parallel arrays, 2M element subarray, M element subarray, and N element subarray, are parallel to one another and spaced $d$ apart. They are used for azimuth and elevation angle calibration, but suffer from estimation failure when the elevation angle is large, as is frequently the case in mobile communication environments.

Another category of 2D coprime antenna array is non-parallel arrays which are further of two types L shape\cite{feng20182}, cross-shaped and V shape antenna array.L shape coprime antenna array is an extension from 1D to 2D arrays where complete coprime arrays can be taken on both the x-axis and y-axis or subarrays can be taken on respective axes of the coordinate plane . Cross-shaped arrays are symmetric arrays that are used to find the azimuth, elevation, and range of a single source in the near-field or far-field. V shape array \cite{elbir2019v} is a generalization of L shape array in which L shape antenna array is tilted by $\theta$ degrees to form V shape but the problem in L and V shape geometry is that it returns coupled estimation results.

The third category of the 2D sparse array is the planar array that can be a coprime planar array, i.e., CPA \cite{zhang2018two} or generalized coprime array GCPA \cite{zheng2017generalized}. In CPA two planes are constructed of M$\times$ M and N$\times$ N where M and N are coprime numbers. In GCPA two rectangular planar subarrays are created with the first subarray as subarray1 having N1 $\times$ M1 sensors and subarray2 having N2 $ \times$ M2 sensors, where N1, N2 are the numbers of sensors on the x-axis and M1, M2 are the numbers of sensors on the y-axis. Both Ni and Mi are coprime numbers. The inter-element spacings among the elements of subarray1 are dx1 = N2 $\cdot \lambda$ / 2 in the x-axis direction and dy1 = M2 $\cdot \lambda$/2 in the y-axis direction and the inter-element spacing among the elements of subarray2 is dx2 = N1 $\cdot \lambda$/2 and dy2 = M1 $ \cdot \lambda$/2, where $\lambda$ represents the wavelength.But CPA and GCPA show poor performance due to a lack of array aperture.

Hence to overcome the problem of phase coupled estimation results and to overcome mutual coupling we propose a planar sparse array which is a phased antenna array. A phased array antenna is used to control the direction of an emitted beam by exploiting constructive interference between two or more radiated signals which is also called beamforming. Although the array-based methods for CPA \cite{zheng2018unfolded} maintain increased DOFs, however, they do not fully exploit the advantage of the non-uniform geometry in terms of high resolution. The 2-D MUSIC algorithm applied in a coprime planar array is feasible, but it requires tremendous spectral search, which is expensive in a real-time environment. Hence, In this paper, we present a novel computationally efficient method for 2-D DOA estimation where the non-uniform property of coprime sparse array is maintained based which not only provides ease in direction of arrival calculation but also helps in attaining a high degree of freedom with less number of holes as compared to other planar array geometries discovered so far.

\section{PRELIMINARIES}

\subsection{Proposed Rectangular Array Geometry}

\begin{figure} 
    \centering
     
    \includegraphics{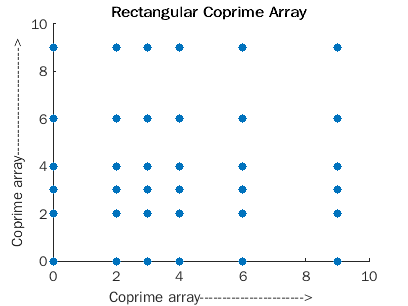}
    \caption{Proposed Rectangular coprime array.}
  \label{fig:coprime1}
\end{figure}

The novel rectangular coprime array (RCPA) is constructed by making a planar geometry of two subarrays, where each subarray consists of a complete conventional coprime array consisting of 2M+N-1 elements where M and N are coprime numbers. In RCPA we take a complete conventional coprime array on both the y-axis and z-axis and then construct a plane from it such that the unique coprime structure of the sparse array is maintained at each point in the lattice, whereas CPA \cite{zheng2019spatial} constructs an M $ \times$ M plane and then N $ \times$ N plane, so it seems to be uniform rectangular array rather than a coprime array in its respective planes. Moreover, in the proposed geometry, there is no need of finding out the superposition of the peaks in the pseudo spectrum, and hence the proposed geometry is computationally efficient. The RCPA is shown in Fig.\ref{fig:coprime1} where N=3, M=2 coprime pair is taken having a total of 2M+N-1=6 elements on each axis thereby creating a plane of (2M+N-1) $ \times $(2M+N-1) elements i.e 6×6=36 elements. First, we create the physical coprime planar array, where a coprime array is given as:

Sensor locations in 1D coprime array are given by the set S with the total number of sensors as T= 2M+N-1 where M and N are coprime numbers and d is the inter-element spacing between the sensors in an array.

\begin{equation}
   S = \{ Mnd \mid 1 \leq  n  \leq N -1 \}  \cup \{ Nmd \mid 0 \leq m \leq 2M -1 \}	 
\end{equation}

For example, lets take M=2, N=3 and, suppose d=1.

\begin{equation}
S = \{ 2n \mid 1 \leq n \leq 2\} \cup \{3m \mid 0 \leq m \leq 3  \}
\end{equation}

\begin{equation}\label{eq4}
S=\{0,2,3,4,6,9\}
\end{equation}

The novel rectangular coprime array (RCPA) is constructed by making a planar geometry of two subarrays, where each subarray consists of a complete conventional coprime array consisting of 2M+N-1 elements where M and N are coprime numbers. In RCPA we take a complete conventional coprime array on both the y-axis and z-axis and then construct a plane from it such that the unique coprime structure of the sparse array is maintained at each point in the lattice, whereas CPA \cite{zheng2019spatial} constructs an M $ \times$ M plane and then N $ \times$ N plane, so it seems to be uniform rectangular array rather than a coprime array in its respective planes. Moreover, in the proposed geometry, there is no need of finding out the superposition of the peaks in the pseudo spectrum, and hence the proposed geometry is computationally efficient. The RCPA is shown in Fig.\ref{fig:coprime1} where N=3, M=2 coprime pair is taken having a total of 2M+N-1=6 elements on each axis thereby creating a plane of (2M+N-1) $ \times $(2M+N-1) elements i.e 6×6=36 elements.

Now we define the set L corresponding to the whose each entry corresponds to the normalized sensor locations in 2D coprime array. It is given by:
\begin{equation}\label{eq5}
L = \{(u, v)\forall u, v \in S \}
\end{equation}

\subsection{2D Array Data Model}

Let M narrowband, uncorrelated, and far-field real sources impinge on the antenna array whose elements are located at $   (u_{x},v_{x}) $ where $(u_{x},v_{x})\in L $ defined in the next section.The $i^{th}$ source have both azimuth and elevation angle as $  \theta, \phi $ respectively from the direction { ( $ \theta_{1} ,\phi_{1} $), ($  \theta_{2} ,\phi_{2} $),($ \theta_{4} ,\phi_{4} $) ..., ( $ \theta_{M} ,\phi_{M } $) }  
with respect to the normal to the array.

The array output can be expressed as

\begin{equation}
Y(t) = \sum_{i=1} ^{M} A(\theta _{i} ^{'},\phi_{i} ^{'}).S_{i} (t) + N(t) 
\label{eq.eq1}
\end{equation}

where $\theta _{i} ^{'},\phi_{i} ^{'} $ are normalized DOA such that $\theta _{i} ^{'}$ = (d/ $\lambda$)sin( $ \theta_{i} $)cos($ \phi_{i}$ ) and $\phi _{i} ^{'}$ = (d/ $\lambda$)sin( $ \theta_{i} $)sin($ \phi_{i}$ ) and d=$\lambda /2$ is the inter sensor spacing.

Each element of the steering vector  A( $ \theta _{i} ^{'},\phi_{i} ^{'} $) = [ $ a(\theta _{1}, \phi_{1}), \cdots , a(\theta _{M}, \phi_{M} $ ] 
corresponding to the sensors at the location $(u_{x},v_{x})\in L $ is  defined as $ e^ {2\pi j \cdot (\theta_{i} ^ {'}\cdot u_{x} + \phi_{i} ^{i}\cdot v_{x})}$ .
Signal $S _{i}(t) = [s_{1}(t), \cdots , s_{M}(t)]^{T} $ is the
source signal vector and N(t) is the white Gaussian noise
vector with zero mean and variance $ \sigma^{2} $.  
t = 1, 2, · · · , T  refers to the sampling time, where T is the total
number of snapshots.

The covariance matrix of Y(t) can be written as:
\begin{equation}\label{eq:2}
R   = E[Y(t)\cdot Y(t) ^{H}]
    = \sum_{i=1} ^{M}\sigma_{i} ^{2}\cdot A(\theta _{i} ^{'},\phi_{i} ^{'})\cdot A^{H} (\theta _{i} ^{'},\phi_{i} ^{'})  
      + \sigma^{2} I
\end{equation}
where $\sigma_{i} ^{2}$ is the $i^{th}$ source power and $\sigma ^{2} $ is the noise power.

Note that the entity $A(\theta _{i} ^{'},\phi_{i} ^{'})\cdot A^{H} (\theta _{i} ^{'},\phi_{i} ^{'})$ in the covariance matrix defined in eq.\ref{eq:2} is of the form $e^{2\pi  (\theta  ^{'},\phi ^{'}) (s_{i}-s_{j})   } $ where $(s_{i}-s_{j})$ $\in$ D i.e the difference coarray having $(s_{i}-s_{j})$ as the difference between $i^{th}$ and $j^{th}$ sensor location.
Applying vectorization operation on eq.\ref{eq:2} and reshaping it we get the autocorrelation vector defined on the difference coarray.

Doa estimation using finite snapshots can be carried out by calculating the finite snapshot version of Y(t) and R as $  Y'(k) $  and $R'$ respectively where $ k= 1,2,3 \cdots K $ 
be K realizations of eq.\ref{eq.eq1}. K are the total number of snapshots. The covariance matrix for which can be estimated as $R'  = \sum_{k=1} ^{K} Y'(k).Y'(k) ^{H} /K $ .Moreover, for sparse array, the covariance matrix R is not hermitian Toeplitz in general but for coarray we can construct a hermitian Toeplitz matrix by the method explained in \cite{liu2016coprime} 
Now partitioning the signal subspace and noise subspace of this matrix and then applying 2D MUSIC on the hermitian Toeplitz matrix will give the desired spectrum of the signal which helps in DOA estimation.

\section{LAGS CALCULATION IN 2D RCPA} 

\begin{figure} 
    \centering
    \includegraphics{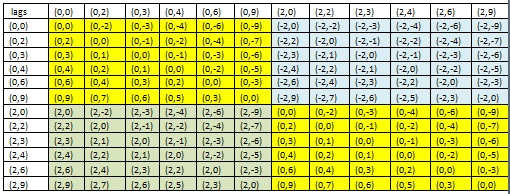}
    \caption{Lag Calculation in 2D RCPA.}
  \label{fig:lag}
\end{figure}

\begin{figure} 
    \centering
    \includegraphics{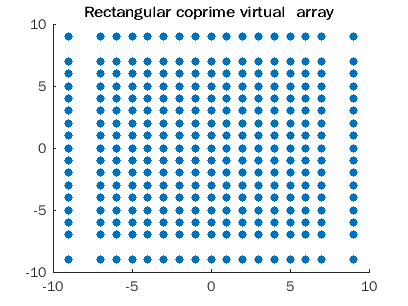}
    \caption{Virtual array of proposed RCPA.}
  \label{fig:coprime2}
\end{figure}

Using a similar concept in a 2D rectangular array \cite{adhikari2019symmetry} we define the virtual co-array by taking the rectangular coordinates of sensors where the spatial covariance between measurements at sensors (i, j, 0) and (p, q, 0) depends solely on the lag (Lx, Ly) = (i - p, j - q) instead of depending on the absolute values of the sensor locations. The lag (Lx, Ly) points to the virtual sensor location (Lx, Ly, 0) which will give the sensor locations in all four quadrants as shown in Fig.\ref{fig:lag}. The lag function is defined as:
 
\begin{equation}
   \begin{cases}
   l(i,j,0)=x(i,j,0) * y(i,j,0) \\
    l(i,j,0)=\sum_{n=0} ^{N-1} \sum_{m=0}^{M-1} x(i,j,0)* y(i,j,0) \\
    where, 
    y(i,j,0)=x(-i,-j,0)   
    \end{cases} 
\end{equation}
\
 where inner sum represents horizontal lags and outer sum corresponds to vertical lags and total lags are contiguous in the range -(MN+M-1)$\leq$i,j $\leq$ (MN+M-1)  as shown in Fig.\ref{fig:coprime2}. The proposed RCPA requires $(3M)^{2} $ total number of sensors to generate the covariance matrix of the dimensions $ (M(N+1))^ {2} \times (M(N+1))^{2} $ whereas in the conventional CPA,  the critical holes in the difference coarray are filled by the proposed holes-filling geometry thus enlarging the contiguous range of the difference coarray and increasing the effective DOF.
 
\begin{figure} 
    \centering
    \includegraphics{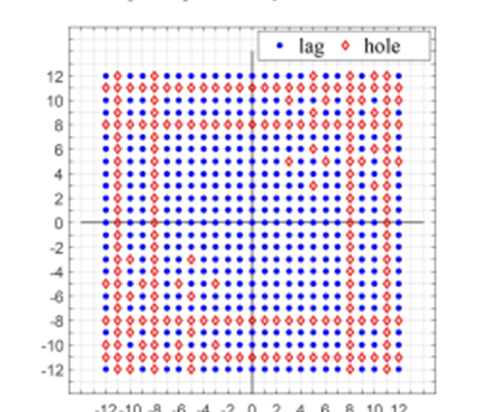}
    \caption{Virtual array of Coprime Planar Array with M1=3, M2=4.}
  \label{fig:hole}
\end{figure}

As shown in Table.\ref{tab:ta} there has been 10.88\% reduction in holes in our proposed geometry as compared to CPA  \cite{yang2020hole} as shown in Fig.\ref{fig:hole}. Moreover, in CPA there is the presence of holes in between the virtual array structure which is also reduced to a far extent in the proposed geometry as we can see the holes are absent only at {i,j= ±8} thereby giving symmetry in the structure of the hole also and this leads to increase the degree of freedom.

\begin{table}
 \caption{Percentage of hole in proposed RCPA}
  \centering
  \begin{tabular}{ll}
    \toprule
    \cmidrule(r){1-2}
    Antenna Array     & Percentage of holes  \\
    \midrule
     CPA & 34.4     \\
    Proposed RCPA & 23.52     \\
  
    \bottomrule
  \end{tabular}
  \label{tab:ta}
\end{table}

\section{ WEIGHT FUNCTION}

\begin{figure} 
    \centering
    \includegraphics{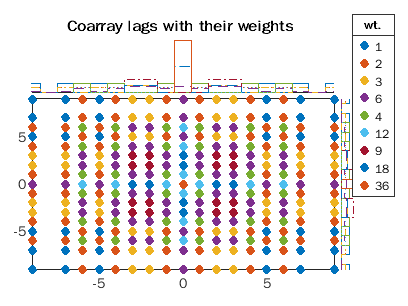}
    \caption{ Virtual array of proposed RCPA along with their weights.}
  \label{fig:coprime33}
\end{figure}

We can observe in the lag matrix defined above in Fig.\ref{fig:lag} that multiple lags are getting repeated in the matrix, which is called redundancy. The weight function can be defined as the count of repetitions of occurrence of virtual sensor positions in the lag matrix, or it can otherwise be considered as the frequency of occurrence of virtual sensors in its coarray. Let g(n) be an integer-valued function that counts the periodicity of $n^{th}$ sensor element in the virtual array $\Phi_{q}$ which can be written as 
\begin{equation}
   \begin{cases}
G_n= \{v_{j1},v_{j2} \cdots v_{jq} \forall \{ v_{j1},v_{j2} \cdots v_{jq} \}   \in   S \\
(v_{j1}+v_{j2} \cdots +v_{jq}) =n   \forall n   \in \Phi_{q}\\
g(n)=|{G_n}| 
\end{cases}
\end{equation}

where $|k|$ is the cardinality of the set k and $v_{jq}$ represents unique lags in a coarray defined by eq.10. So counting the weights of all such unique virtual sensors we can plot their frequency with the colors indicating the virtual sensors with similar frequency of occurence by one color. Co-array lags of the proposed RCPA are shown in Fig.\ref{fig:coprime33}. along with their weights.

\section{ SIMULATION RESULTS}

\begin{figure} 
    \centering
    \includegraphics{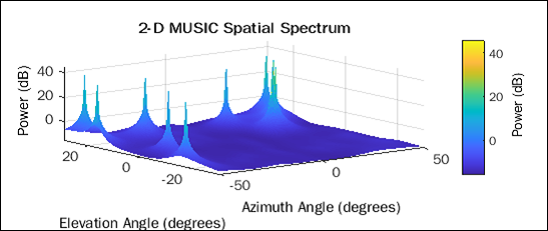}
    \caption{ 2D Music Spectrum of RCPA}
  \label{fig:musicspectrumrcpa}
\end{figure}

2D ( azimuth and elevation) angle calculations successfully overcome the physical resolution which is generally provided by the antenna aperture. 2D MUSIC algorithm is applied to the proposed RCPA geometry. Fig.\ref{fig:musicspectrumrcpa} gives the 2D MUSIC spectrum of the virtual array when 8 signals impinges upon it having azimuth angle in the span of -50$^{\circ}$ to 50$^{\circ}$ and elevation angle in the span of-90$^{\circ}$  to 90$^{\circ}$ with the step of 0.5$^{\circ}$ .

\subsection{RMSE with changing SNR }
\begin{figure} 
    \centering
    \includegraphics{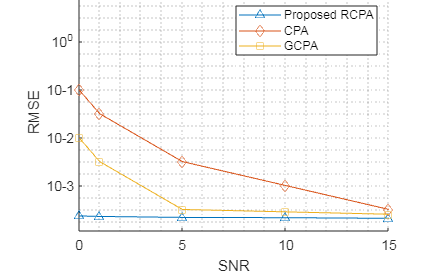}
    \caption{ RMSE versus SNR(dB)}
  \label{fig:rmse1}
\end{figure}

RMSE (root mean square error) for the DOA can be defined as

\begin{equation}
    RMSE=\sqrt{1/(LQ)\sum_{i=1}^{L}\sum_{j=1}^{Q}\{(\theta_{q}(i)-\theta_{q})^2 + (\phi_{q}(i)-\phi_{q})^2 \}}
\end{equation}

where $\theta_{q}(i)$ and $\phi_{q}(i)$ are the estimates of $\theta_{q}$ and $\phi_{q}$ for the $i^{th} $ Monte Carlo trial, i = 1, . . ., L. In the first set of simulation, we consider the case where we change the SNR and see the effect of increasing SNR on the proposed geometry. To enable a feasible comparison we consider a planar array composed of a coprime pair as (M=2, N=3). Let us take assume 49 uncorrelated sources falling on it whose normalized DOA are picked up randomly from ( $\theta^{'} ,\phi^{'}) $ $ =$  $[-0.5,0.5]$. Let the number of snapshots is fixed at 500 and SNR varied from 0dB to 15dB. We use L = 500 independent trials in the Monte carlo simulations. Fig.\ref{fig:rmse1} shows the RMSE performance comparison among the proposed geometry and existing planar geometries like CPA and GCPA(Generalized coprime array) as a function of input signal-to-noise ratio (SNR), where RMSE between azimuth and elevation angle decreases as signal to noise ratio increases. It can be observed that RMSE of proposed geometry is of the order of $10^{-3}$ which shows significant improvement in DOA estimation..

\subsection{RMSE with changing Snapshot.}
\begin{figure} 
    \centering
    \includegraphics{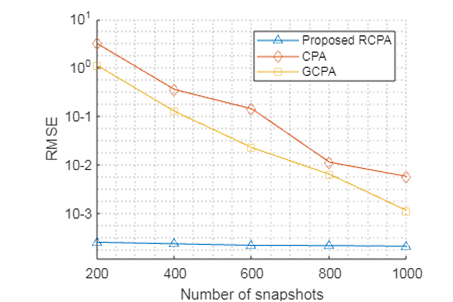}
    \caption{ RMSE versus number of snapshots}
  \label{fig:rmse2}
\end{figure}

Considering same virtual planar array composed of a coprime pair as (M=2, N=3). Let us take assume 49 uncorrelated sources falling on it whose normalized DOA are picked up randomly from ( $\theta^{'} ,\phi^{'}) $ $ =$  $[-0.5,0.5]$. Let the number of SNR is fixed to 0dB and the number of snapshots varies from 200 to 1000. Fig.\ref{fig:rmse2}, shows the RMSE performance comparison among the proposed geometry and existing planar geometries like CPA and GCPA(Generalized coprime array) as a function of number of snapshots, where RMSE between azimuth and elevation angle decreases as number of snapshots increases. It can be observed that RMSE of proposed geometry is of the order of $10^{-3}$ which shows significant improvement in DOA estimation.

\section{Optimizaton}
\begin{figure} 
    \centering
    \includegraphics{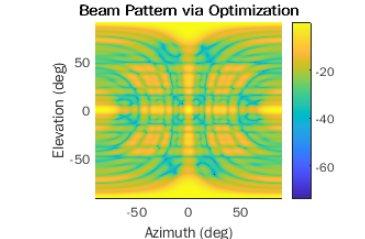}
    \caption{ Optimized Beam pattern}
  \label{fig:Picture1}
\end{figure}

Optimization is used to derive the pattern synthesis weights. We tried to optimize the pattern whose main lobe is along azimuth and elevation 0 degrees. The pattern should satisfy the following constraints like Maximize the directivity, Suppress interferences 30 dB below main lobe, side lobe levels should be 17 dB below main lobe and within -20 and 20 degrees azimuth or elevation. To improve directivity needs to minimize the total radiated power, which is given by w'*Rn*w which is our objective function.
A second-order cone programming solver is used to derive the array weights that can provide us with the desired pattern as shown in Fig.\ref{fig:Picture1}. The table shows that we can meet the calculated requirements and the weight computation time is 13.99 seconds.

\begin{table}
 \caption{Performance Metrics}
  \centering
  \begin{tabular}{lll}
    \toprule
    \cmidrule(r){1-2}
    Metric     & Measured & Calculated \\
    \midrule
     Directivity(dBi)&[14.8077]& [14.8066]    \\
    Interference Suppression(dB)&[-29.3044, -40.0438]&[-30, -40]    \\
  Sidelobe Over requirement percentage & [0] & [0] \\
    \bottomrule
  \end{tabular}
  \label{tab:tab1}
\end{table}

\section{ Conclusion}

In this paper, we have proposed a new class of rectangular arrays which is a combination of sensors distributed over a plane thereby maintaining the uniqueness of a coprime array, whose difference co-array generates a much wider spectrum giving rise to the large fully augmentable range. This proposed planar array geometry leads to a more flexible array layout which leads to significant reduction in number of holes i.e.$ 10\% $ as compared to CPA whose virtual array shows holes within the contiguous range also. The optimized RCPA geometry shows significant performance improvement in terms of achieving directivity and interference supression in the array look-up direction which is used in beam steering. The advantage of the proposed antenna array geometry is that it provides automatically paired 2D DOA angles of the signals and attains better performance due to its higher DOF and larger array aperture.


\bibliographystyle{unsrt}  
\bibliography{references}  


\end{document}